\documentclass[twocolumn,numberedappendix,twocolappendix,appendixfloats]{openjournal_dhayaa}

\usepackage{amsmath}
\usepackage{empheq} 
\usepackage{color}
\usepackage{xcolor, multirow}

\usepackage{natbib}
\setcitestyle{aysep={}}

\usepackage[colorlinks=true
  ,urlcolor=blue
  ,anchorcolor=blue
  ,citecolor=blue
  ,filecolor=blue
  ,linkcolor=blue
  ,menucolor=blue
  ,linktocpage=true
  ,pdfproducer=medialab
  ,pdfa=true
]{hyperref}

\usepackage{etoolbox}
\makeatletter

\usepackage{graphicx}	
\usepackage{amsmath}	
\usepackage{color}
\usepackage{xcolor}

\newcommand{\OrcidID}[1]{ \href[urlcolor = red]{https://orcid.org/#1}{\textcolor{lightgray}{\faOrcid}}}
\newcommand{\OrcidIDName}[2]{\href{https://orcid.org/#1}{#2}}

\defcitealias{Pandey2024godmax}{P24}
\defcitealias{Schneider2019Baryonification}{S19}

\begin{document}
\allowdisplaybreaks

\title[Cosmic structures in Ricci-inverse gravity]{Cosmic structures in Ricci-inverse theories of gravity}

\author{\OrcidIDName{0000-0002-2795-5929}{Scomparin Mattia}$^{1}$}
\email{$^{\star}$mattia.scompa@gmail.com}
\affiliation{$^{1}$ Mogliano Veneto, 31021, TV, Italy}

\begin{abstract}
We discuss a no-go theorem for the novel Ricci-inverse theory of modified gravity. By considering a static spherically symmetric matter distribution embedded within a de Sitter cosmology, we demonstrate that achieving a stable Sub-Horizon non-relativistic Weak-Field limit is unattainable in any of the models previously proposed to mitigate certain cosmological and inflationary instabilities.  We explore potential strategies to address this challenge, suggesting a novel methodology for constructing stable models that adhere to the Sub-Horizon non-relativistic Weak-Field limit. These models are shown to maintain full consistency with the predictions of General Relativity at small scales.
\end{abstract}


\section{Introduction}
\label{sec:intro}

The modeling of the current accelerated expansion of the Universe constitutes one of the biggest challenges of modern cosmology (\cite{f,g,h}). More specifically, the notable cosmological difficulties linked to the standard $\Lambda$CDM model (\cite{l,m,r}) have led to a surge of interest in Modified Gravity (MG) theories, which are considered as potential alternatives to General Relativity (GR). 

A novel class of fourth-order MG models is represented by the so-called \textit{Ricci-inverse} gravity (\cite{a}). In this framework, the Einstein-Hilbert action is extended through the inclusion of a function $f(R,A)$, which depends on the Ricci scalar $R$ and the \textit{anticurvature} scalar $A$, the latter being defined as the trace of the \textit{Ricci-inverse tensor} $A^{\mu\nu}$. In particular
\begin{equation}\label{eq:def}
A^{\mu\sigma}R_{\sigma\nu}=\delta^\mu\!_\nu\,.
\end{equation}

The Ricci-inverse theory has been shown to encounter both cosmological and inflationary no-go theorems. In particular, actions that incorporate terms that are linear in any positive or negative power of $A$ are ruled out as potential candidates for dark energy (\cite{a}). Furthermore, it is not possible to attain stable isotropic inflation through any linear combination of $R$, $A$, and $A^2$ (\cite{d,i}).

In order to circumvent the cosmological no-go theorem presented in (\cite{a}), an initial strategy proposes the incorporation of simple non-linear terms. However, a thorough investigation is necessary, as typical non-linear combinations of $R$ and $A$ may lead to the emergence of ghosts or other types of instability (\cite{n,s}).

In light of such considerations, this study aims to investigate a third no-go theorem that pertains to the stability and consistency of the Ricci-inverse theory in relation to the predictions of General Relativity (GR) at small scales. Focusing on a static spherically symmetric matter distribution within a de Sitter cosmology, our key findings are as follows: (i) it is not possible to achieve a stable Sub-Horizon non-relativistic Weak-Field limit through any linear combination of $R$ with $A$ and $A^2$ (\cite{d}), nor through any non-linear terms suggested in (\cite{a}) to circumvent the cosmological no-go theorem; (ii) we have identified specific non-linear combinations of $A$ and $R$ that effectively prevent instabilities from the Sub-Horizon non-relativistic Weak-Field perspective; (iii) when stability is guaranteed, this combination is fully consistent with the predictions of GR, thereby demonstrating the difficulty of detecting signatures of Ricci-inverse theories at small scales using astrophysical objects such as stars and galaxy clusters (\cite{b,o,p}).

The organization of this paper is structured to enhance the clarity of our findings. In Section \ref{sec:RI}, we provide a succinct summary of the entire Ricci-inverse theory along with the relevant covariant field equations. Section \ref{sec:dS} focuses on the de Sitter background, while Section \ref{sec:sph} is dedicated to deriving the perturbed equations related to a static spherically symmetric matter source. Following this, we investigate the corresponding Weak-Field limit. In Section \ref{sec:nogo}, we establish that a general no-go theorem effectively declares the occurrence of divergences and ghosts in any linear combination of $R$ with $A$ and $A^2$, thus ruling it out as a plausible cosmological candidate. Section \ref{sec:circnogo} explores possible strategies to circumvent our no-go theorem, emphasizing a stable action that is consistent with the predictions of General Relativity. Finally, Section \ref{sec:conc} summarizes our conclusions.

We utilize the metric signature $(-, +, +, +)$ and set the reduced Planck mass to unity. Greek indices are used to denote values ranging from 0 to 3.


\section{The Ricci-inverse theory}
\label{sec:RI}
Let us consider the full action for the Ricci-inverse theory of gravity (\cite{a})
\begin{equation}\label{eq:action}
S=\int d^4x\sqrt{-g}\,\Big[f(R,A)+\mathcal{L}_m\Big],
\end{equation}
where $g$ is the determinant of the metric $g_{\mu\nu}$, and $\mathcal{L}_m$ is the matter Lagrangian, that we assume coupled with the metric only. The arbitrary function $f(R,A)$ depends on the \textit{Ricci scalar} $R$ and the \textit{anticurvature scalar} $A\equiv g_{\mu\nu}A^{\mu\nu}$.

By differentiating Eq. \eqref{eq:action}, the covariant equation of motion  with respect to the metric field $g_{\mu\nu}$ is $\delta S/\delta g_{\mu\nu} = 0$,  whose explicit expression is (\cite{a})
\begin{equation}\label{eq:covEOM}
\mathcal{G}^{\mu\nu}=T^{\mu\nu}\,.
\end{equation}
We introduced the modified Einstein tensor 
\begin{align}
\mathcal{G}^{\mu\nu}&\equiv\partial_Rf R^{\mu\nu}
- \tfrac{1}{2} f g^{\mu\nu}- \partial_Af A^{\mu\nu}+ g^{\mu\nu}\nabla^{\alpha}{\nabla_{\alpha}{\,\partial_Rf}} \nonumber\\
&- \tfrac{1}{2} \nabla^{\alpha}{\nabla_{\alpha}{(\partial_A f A_{\sigma}^{\mu}A^{\nu\sigma}})}
+g^{\rho\mu} (\nabla_{\alpha}{\nabla_{\rho}{\,\partial_A f}}) A^{\alpha}_{\sigma} A^{\nu\sigma}
\nonumber\\
&- \tfrac{1}{2} g^{\mu\nu} \nabla_{\alpha}{\nabla_{\beta}{(\partial_Af A_{\sigma}^{\alpha} A^{\beta\sigma})}}
- \nabla^{\mu}{\nabla^{\nu}{\partial_Rf}} \,,
\end{align}
and the energy-momentum tensor $T_{\mu\nu}$ defined as
\begin{equation}\label{eq:T}
T^{\mu\nu}\equiv\frac{2}{\sqrt{-g}}\frac{\delta(\sqrt{-g}\mathcal{L}_m)}{\delta g_{\mu\nu}}\,,
\end{equation}
In our notation the $\nabla_\mu$ symbol denotes the covariant derivative, whereas
the subscripts $A$ and $R$ stand for partial derivatives, e.g. $f_{AAR}\equiv\partial_R\partial_A\partial_A f$.


\section{Ricci-inverse in de Sitter background}
\label{sec:dS}

Consider a spatially flat de Sitter cosmological background. Assuming the Friedmann Lemaitre Robertson Walker (FLRW) coordinates $(\tau, \rho, \theta, \phi)$, the metric can be written as
\begin{equation}\label{eq:dSmetric}
ds_{(0)}^2=-d\tau^2+e^{2H\tau}\big(d\rho^2+\rho^2d\Omega_2^2\big)\,,
\end{equation}
where $H$ is the constant Hubble expansion rate and $d\Omega^2_2$ the solid angle-element.

Using \eqref{eq:dSmetric}, the resulting background Ricci scalar $R^{(0)}$ and background anticurvature scalar $A^{(0)}$ are, respectively
\begin{equation}\label{eq:backRA}
R^{(0)}=12H^{2}\,,\qquad A^{(0)}=\tfrac{4}{3}H^{-2}\,.
\end{equation}
We take the trace of the equation of motion \eqref{eq:covEOM}, and using \eqref{eq:backRA} we finally get, in vacuum  (\cite{a})
\begin{equation}\label{eq:dsEOM}
18 f_R^{(0)} H^2 - 2 H^{-2} f_A^{(0)} - 3 f^{(0)} = 0\,,
\end{equation}
where we introduced the notation $f_{\cdot}^{(0)}\equiv f_{\cdot}\rvert_{R^{(0)},A^{(0)}}$ to indicate evaluation with respect to background quantities.

The subsequent sections will examine cosmic structures represented in spherical Schwarzschild-like coordinates $(t, r, \theta, \phi)$, which can be derived from the FLRW coordinates through the transformation outlined in \cite{b}. This transformation is given by the following equations:
\begin{subequations}
\label{eq:tr}
\begin{empheq}[left=\empheqlbrace]{align}
\label{eq:tr1}
\, &\tau(t,r) = t + \frac{1}{2H}\ln\big(1-H^2r^2\big)\,,\\
\label{eq:tr2}
\,&\rho(t,r)=\frac{re^{-Ht}}{\sqrt{1-H^2r^2}}\,,
\end{empheq}
\end{subequations}
with the condition that $1-H^2r^2\ge0$. 

By expressing the metric \eqref{eq:dSmetric} in the Schwarzschild-like coordinates \eqref{eq:tr}, it can be readily observed that the de Sitter background can be reformulated as follows:
\begin{equation}\label{eq:dSmetricSW}
ds_{(0)}^2=-\big(1-H^2r^2\big)dt^2+\frac{dr^2}{1-H^2r^2}+r^2d\Omega_2^2\,.
\end{equation}


\section{Static spherically symmetric matter distribution in Ricci-inverse gravity}
\label{sec:sph}

Let us embed a static and spherically symmetric structure into the de Sitter cosmological background \eqref{eq:dSmetric}.  This source influences the surrounding spacetime, which, when expressed in spherical Schwarzschild-like coordinates, takes the form
\begin{equation}\label{eq:sw}
ds^2=-e^{\nu(r)}dt^2+e^{\lambda(r)}dr^2+r^2d\Omega^2_2\,,
\end{equation}
where $\nu(r)$ and $\lambda(r)$ represent two metric potentials that depend on the radial coordinate.

Writing down $R$ and $A$ in terms of the spherical metric \eqref{eq:sw}, it is easy to find that
\begin{subequations}
\begin{empheq}[]{align}
	\, &R = \Big(\xi_1+r \xi_2-r \xi_3\Big)\, e^{-\lambda} {r}^{-2}\,,\\	
	\,&A = \Big(4r \xi_1^{-1}+ \xi_2^{-1}- \xi_3^{-1}\Big)\,r e^\lambda\,,
\end{empheq}
\end{subequations}
where 
\begin{subequations}
\begin{empheq}[]{align}
	\, &\xi_1 \equiv r (\lambda'-\nu' )+ 2(e^{\lambda}-1)\,,\\	
	\, &\xi_2 \equiv \lambda'- \tfrac{1}{4}(  \nu'^2 -   \nu' \lambda' +2 \nu'')\,,\\	
	\,&\xi_3 \equiv \nu'+ \tfrac{1}{4}( \nu'^2 -   \nu'\lambda' + 2 \nu'')\,,
\end{empheq}
\end{subequations}
In our notation, the symbol $\,\!'$ represents the derivative with respect to the radial coordinate $r$. It is evident that $A$ becomes singular when any of the variables $\xi_1$, $\xi_2$, or $\xi_3$ equal zero. This indicates that if a solution encounters any of these conditions, it results in a singularity that invalidates the model.

In the context of a spherically symmetric perfect fluid model characterized by an energy density $\varepsilon(r)$ and pressure $P(r)$, the energy-momentum tensor can be expressed as follows:
\begin{equation}\label{eq:Tsph}
T^{\mu}\!_\nu\equiv \mbox{diag}\Big\{-\varepsilon(r),P(r),P(r),P(r)\Big\}\,.
\end{equation}

By substituting the explicit expressions from equations \eqref{eq:sw} and \eqref{eq:Tsph} into equation \eqref{eq:covEOM}, it can be determined that the pertinent equations of motion correspond to the $t$-$t$ and $\theta$-$\theta$ components. These components are expressed as linear combinations of the derivatives of the function $f$ (specifically, $f_i = f, f_A, f_R, \ldots$), which are further multiplied by polynomials in the variable $r$ and derivatives of the metric potentials, denoted as $\mathcal{P}_i$ and $\mathcal{Q}_i$:
\begin{subequations}
\label{eq:eomsph}
\begin{empheq}[left=\empheqlbrace]{align}
\label{eq:eomsph1}
\,&\varepsilon\, e^{-\nu}=\sum f_i\mathcal{P}_i\,,\\
\label{eq:eomsph2}
\,&Pr^{-2}=\sum f_i\mathcal{Q}_i\,.
\end{empheq}
\end{subequations}
The lengthy expressions for $\mathcal{P}_i$ and $\mathcal{Q}_i$ are omitted for brevity.


\subsection{Sub-Horizon non-relativistic Weak-Field limit}
\label{sec:weak}

The alignment of Modified Gravity (MG) theories with the predictions of General Relativity (GR) can be evaluated on small scales by examining the Sub-Horizon non-relativistic Weak-Field limit. To initiate this analysis, we can perturb the metric potentials around their cosmological values as follows:
\begin{equation}
\nu(r)\sim\nu^{(0)}(r)+\delta\nu(r)\,,\qquad \lambda(r)\sim\lambda^{(0)}(r)+\delta\lambda(r)\,,
\end{equation}
where it is assumed that $\delta\nu\ll \nu^{(0)}$ and $\delta\lambda\ll \lambda^{(0)}$. As the radial coordinate $r$ approaches the de Sitter horizon, both $\delta\lambda$ and $\delta\nu$ tend to zero, resulting in the predominance of the background de Sitter metric \eqref{eq:dSmetricSW}.

In light of such decompositions, the Ricci scalar and the anticurvature scalar can be expressed in the following manner:
\begin{equation}
R \sim R^{(0)} + \delta R\,, \qquad A \sim A^{(0)} + \delta A\,,
\end{equation}
where the background quantities \( R^{(0)} \) and \( A^{(0)} \) are defined in Equation \eqref{eq:backRA}. The perturbations are given by:
\begin{align}
\delta R &= \delta \nu'' (H^2 r^2 - 1) + \delta \nu' \frac{5H^2 r^2 - 2}{r} \nonumber\\
&- \delta \lambda' \frac{3H^2 r^2 - 2}{r} - 2\delta \lambda \frac{6H^2 r^2 - 1}{r^2}\,,
\end{align}
and
\begin{align}
\delta A &= -\frac{1}{9} \delta \nu'' \frac{H^2 r^2 - 1}{H^4} - \frac{1}{9} \delta \nu' \frac{5H^2 r^2 - 2}{H^4 r} \nonumber\\
&+ \frac{1}{9} \delta \lambda' \frac{3H^2 r^2 - 2}{H^4 r} + \frac{2}{9} \delta \lambda \frac{6H^2 r^2 - 1}{H^4 r^2}\,.
\end{align}

As a result, the scalar function \( f(R, A) \) can be decomposed as \( f \sim f^{(0)} + \delta f \), where
\begin{equation}\label{eq:decF}
\delta f = f^{(0)}_R \delta R + f^{(0)}_A \delta A\,.
\end{equation}
This methodology can similarly be applied to any derivative of the function \( f \).

Assuming a mass distribution $M(r)$ of the matter source  defined by
\begin{equation}
M(r)\equiv4\pi \int_0^r\!\!\!s^2\varepsilon(s) \,ds\,,\qquad M(r\rightarrow+\infty)\equiv\mathcal{M}\,,
\end{equation}
the Sub-Horizon non-relativistic Weak-Field limit is systematically approached by sequentially addressing: (i) the Weak-Field condition $\delta\nu' \sim \delta\lambda\sim M/r \ll 1$, (ii) the sub-horizon condition $x\equiv Hr \ll 1$, and (iii) the non-relativistic Newtonian condition $P\ll\varepsilon$. Under these conditions, the metric potentials $\delta\nu$ and $\delta\lambda$ can be expressed in relation to the \textit{Newtonian potential} and \textit{curvature perturbations} as follows:
\begin{equation}\label{eq:metricPot}
\Phi'(r) = \frac{\delta\nu'(r)}{2}\,,\qquad\Psi'(r)=\frac{\delta\lambda(r)}{2r}\,.
\end{equation}
In the context of General Relativity, it is well established that 
\begin{equation}
\label{eq:metricPotA}
\Phi'_{GR}=\Psi'_{GR}=M/r^{2}\,.
\end{equation}

\begin{widetext}
We will now implement the procedure within the context of our theoretical framework. Beginning with the Weak-Field condition, and following extensive calculations, the field equations \eqref{eq:eomsph} can be reformulated as
\begin{align}\label{eq:pert1}
\varepsilon {(1-{x}^{2})}^{-1}  &=  \Big\{3{x}^{4} f^{(0)}_R - \tfrac{1}{3}{r}^{4} f^{(0)}_{A} - \tfrac{1}{2}{r}^{2} f^{(0)} {x}^{2}\Big\} {\Big\{{r}^{2} ({x}^{2}-1) {x}^{2}\Big\}}^{-1}\nonumber\\
&+\tfrac{1}{81}\delta\nu'' \Big\{5{r}^{8} (5-6{x}^{2}) f^{(0)}_{AA}-3{r}^{6} (37{x}^{2}-27) {x}^{2} f^{(0)}_{A}-54{r}^{4} ({x}^{2}-1) {x}^{4} f^{(0)}_{AR}+405(2{x}^{2}-1) {x}^{8} f^{(0)}_{RR}\Big\} {\Big\{{r}^{2} ({x}^{2}-1) {x}^{6}\Big\}}^{-1}\nonumber\\
&+\tfrac{1}{81}\delta\nu' \Big\{2{r}^{8} f^{(0)}_{AA}-9{r}^{6} {x}^{4} f^{(0)}_{A}-54{r}^{4} (5{x}^{2}-2) {x}^{4} f^{(0)}_{AR}+324(5{x}^{2}-1) {x}^{8} f^{(0)}_{RR}\Big\} {\Big\{{r}^{3} ({x}^{2}-1) {x}^{6}\Big\}}^{-1}\nonumber\\
&-\tfrac{1}{81}\delta\lambda' \Big\{4{r}^{8} (-9{x}^{4}+5{x}^{2}+1) f^{(0)}_{AA}+3{r}^{6} (-40{x}^{4}+17{x}^{2}+2) {x}^{2} f^{(0)}_{A}-54{r}^{4} (3{x}^{2}-2) {x}^{6} f^{(0)}_{AR}\nonumber\\
&\qquad\qquad+81{r}^{2} ({x}^{2}-1) {x}^{8} f^{(0)}_R-486(4{x}^{2}-1) {x}^{10} f^{(0)}_{RR}\Big\} {\Big\{{r}^{3} ({x}^{2}-1) {x}^{8}\Big\}}^{-1}\nonumber\\
&+\tfrac{1}{81}\delta{\lambda} \Big\{2{r}^{8} (-18{x}^{4}+3{x}^{2}+2) f^{(0)}_{AA}+3{r}^{6} (-27{x}^{4}+5{x}^{2}+2) {x}^{2} f^{(0)}_{A}+108{r}^{4} (6{x}^{2}-1) {x}^{6} f^{(0)}_{AR}\nonumber\\
&\qquad\qquad+81{r}^{2} (3{x}^{2}-1) {x}^{8} f^{(0)}_R-162(18{x}^{2}-1) {x}^{10} f^{(0)}_{RR}\Big\} {\Big\{{r}^{4} ({x}^{2}-1) {x}^{8}\Big\}}^{-1}\nonumber\\
&+\tfrac{1}{6}\delta{\nu} \Big\{2{r}^{4} f^{(0)}_{A}+3{r}^{2} f^{(0)} {x}^{2}-18{x}^{4} f^{(0)}_R\Big\} {\Big\{{r}^{2} ({x}^{2}-1) {x}^{2}\Big\}}^{-1}\nonumber\\
&+\tfrac{1}{81}\delta\nu''' \Big\{-4{r}^{8} (3{x}^{2}-1) f^{(0)}_{AA}-3{r}^{6} (13{x}^{2}-4) {x}^{2} f^{(0)}_{A}+81{x}^{10} f^{(0)}_{RR}\Big\}r^{-1}{x}^{-8}\nonumber\\
&+\tfrac{1}{81}\delta\lambda'' \Big\{{r}^{8} (27{x}^{4}-28{x}^{2}+2) f^{(0)}_{AA}+3{r}^{6} (21{x}^{4}-20{x}^{2}+1) {x}^{2} f^{(0)}_{A}-81(3{x}^{2}-2) {x}^{10} f^{(0)}_{RR}\Big\} {\Big\{{r}^{2} ({x}^{2}-1) {x}^{8}\Big\}}^{-1}\nonumber\\
&- \tfrac{1}{81}\delta\nu'''' {r}^{6} \Big\{({r}^{2} f^{(0)}_{AA}+3{x}^{2} f^{(0)}_{A}) ({x}^{2}-1)\Big\} {x}^{-8}\nonumber\\
&+\tfrac{1}{81}\delta\lambda''' {r}^{5} \Big\{{r}^{2} (3{x}^{2}-2) f^{(0)}_{AA}+3(2{x}^{2}-1) {x}^{2} f^{(0)}_{A}\Big\} {x}^{-8} \,,
\end{align}
and
\begin{align}\label{eq:pert2}
Pr^{-2} &= - \tfrac{1}{3}f_{A} {x}^{-2} - \tfrac{1}{2}f {r}^{-2}+3{x}^{2} f^{(0)}_R {r}^{-4}\nonumber\\
&+\tfrac{1}{162}\delta\nu'' \Big\{4{r}^{8} (-15{x}^{4}+9{x}^{2}+1) f^{(0)}_{AA}-3{r}^{6} (45{x}^{4}-37{x}^{2}+2) {x}^{2} f^{(0)}_{A}-108{r}^{4} ({x}^{2}-1) {x}^{6} f^{(0)}_{AR}\nonumber\\
&\qquad\qquad-81{r}^{2} ({x}^{2}-1) {x}^{8} f^{(0)}_R+324(5{x}^{4}-6{x}^{2}+1) {x}^{8} f^{(0)}_{RR}\Big\} {r}^{-4} {x}^{-8}\nonumber\\
&- \tfrac{1}{162}\delta\nu' \Big\{2{r}^{8} (3{x}^{2}+2) f^{(0)}_{AA}+3{r}^{6} (10{x}^{4}+{x}^{2}-2) {x}^{2} f^{(0)}_{A}+108{r}^{4} (5{x}^{2}-2) {x}^{6} f^{(0)}_{AR}\nonumber\\
&\qquad\qquad+81{r}^{2} (4{x}^{2}-1) {x}^{8} f^{(0)}_R+162(-20{x}^{4}+9{x}^{2}+2) {x}^{8} f^{(0)}_{RR}\Big\} {r}^{-5} {x}^{-8}\nonumber\\
&-\tfrac{1}{162}\delta\lambda' \Big\{2{r}^{8} (-36{x}^{4}+5{x}^{2}+4) f^{(0)}_{AA}+3{r}^{6} (-80{x}^{4}+35{x}^{2}+6) {x}^{2} f^{(0)}_{A}\nonumber\\
&\qquad\qquad-108{r}^{4} (3{x}^{2}-2) {x}^{6} f^{(0)}_{AR}-81{r}^{2} (2{x}^{2}-1) {x}^{8} f^{(0)}_R+486(8{x}^{2}-7) {x}^{10} f^{(0)}_{RR}\Big\} {r}^{-5} {x}^{-8}\nonumber\\
&+\tfrac{1}{81}\delta{\lambda} \Big\{2{r}^{8} (-18{x}^{4}+3{x}^{2}+4) f^{(0)}_{AA}+3{r}^{6} (-27{x}^{4}+12{x}^{2}+2) {x}^{2} f^{(0)}_{A}+108{r}^{4} (6{x}^{2}-1) {x}^{6} f^{(0)}_{AR}\nonumber\\
&\qquad\qquad+243{r}^{2} {x}^{10} f^{(0)}_R+162(-18{x}^{4}+{x}^{2}+2) {x}^{8} f^{(0)}_{RR}\Big\} {r}^{-6} {x}^{-8}\nonumber\\
&+\tfrac{1}{54}\delta\nu''' \Big\{-2{r}^{8} (4{x}^{4}-5{x}^{2}+1) f^{(0)}_{AA}-{r}^{6} (14{x}^{4}-19{x}^{2}+5) {x}^{2} f^{(0)}_{A}+54({x}^{4}-2{x}^{2}+1) {x}^{8} f^{(0)}_{RR}\Big\} {r}^{-3} {x}^{-8}\nonumber\\
&+\tfrac{1}{81}\delta\lambda'' \Big\{{r}^{8} (27{x}^{2}-25) f^{(0)}_{AA}+3{r}^{6} (21{x}^{4}-22{x}^{2}+2) f^{(0)}_{A}-81(3{x}^{4}-5{x}^{2}+2) {x}^{6} f^{(0)}_{RR}\Big\} {r}^{-4} {x}^{-6}\nonumber\\
&- \tfrac{1}{162}\delta\nu'''' {r}^{4} \Big\{(2{r}^{2} f^{(0)}_{AA}+3{x}^{2} f^{(0)}_{A}) ({x}^{4}-2{x}^{2}+1)\Big\} {x}^{-8}\nonumber\\
&+\tfrac{1}{162}\delta\lambda''' {r}^{3} \Big\{2{r}^{2} (3{x}^{4}-5{x}^{2}+2) f^{(0)}_{AA}+3(4{x}^{4}-7{x}^{2}+3) {x}^{2} f^{(0)}_{A}\Big\} {x}^{-8} \,.
\end{align}
\end{widetext}

The parametrization of \( f(R,A) \) leads to the emergence of two distinct types of instabilities in Eqs. \eqref{eq:pert1} and \eqref{eq:pert2}. These instabilities are characterized by (i) the occurrence of divergences when the Sub-Horizon limit \( x \rightarrow 0 \) is applied, and (ii) the presence of ghost instabilities arising from terms that involve higher-order derivatives of \( \delta\nu \) and \( \delta\lambda \).

To mitigate these issues, one potential approach is to identify a finely-tuned set of \( f(R,A) \) functions that could eliminate both divergences and ghost instabilities.
However, prior to exploring potential solutions, it is essential to establish that achieving a stable Sub-Horizon non-relativistic Weak-Field limit is unattainable in any linear combination of \( R \) with \( A \) and \( A^2 \) (as noted in \cite{d}), or in any non-linear terms suggested to bypass the cosmological no-go theorem outlined in \cite{a}.


\section{A no-go theorem}
\label{sec:nogo}

In this section, we examine the straightforward scenario represented by the equation
\begin{equation}\label{eq:case1:f}
f(R,A)=R+kA+\ell A^2\,,
\end{equation}
where \( k \) and \( \ell \) are constants. By solving the background equation \eqref{eq:dsEOM} with respect to the parameter \( k \), we obtain the following expression:
\begin{equation}\label{eq:case1:k}
k = - \frac{16\ell {r}^{6}+27{x}^{6}}{9{r}^{4} {x}^{2}}\,.
\end{equation}
We  replace the aforementioned parametrization \eqref{eq:case1:f} into the simplified equations \eqref{eq:pert1} and \eqref{eq:pert2}. 
\begin{widetext}
By applying the background relation \eqref{eq:case1:k}, the following results are obtained.
\begin{align} \label{eq:case1:eom1}
\varepsilon {(1-{x}^{2})}^{-1} &=- \tfrac{1}{243}\delta\nu'' \Big\{476\ell {r}^{6} {x}^{2}-366\ell {r}^{6}-999{x}^{8}+729{x}^{6}\Big\} {\Big\{(x^2-1) {x}^{6}\Big\}}^{-1}\nonumber\\
& - \tfrac{1}{81}\delta\nu' \Big\{8\ell {r}^{6} {x}^{2}-4\ell {r}^{6}-27{x}^{8}\Big\} {\Big\{r (x^2-1) {x}^{6}\Big\}}^{-1}\nonumber\\
&+\tfrac{1}{243}\delta\lambda' \Big\{536\ell {r}^{6} {x}^{4}-256\ell {r}^{6} {x}^{2}-40\ell {r}^{6}-837{x}^{10}+216{x}^{8}+54{x}^{6}\Big\} {\Big\{r (x^2-1) {x}^{8}\Big\}}^{-1}\nonumber\\
& - \tfrac{2}{243}\delta\lambda \Big\{216\ell {r}^{6} {x}^{4}-38\ell {r}^{6} {x}^{2}-20\ell {r}^{6}-729{x}^{10}+189{x}^{8}+27{x}^{6}\Big\} {\Big\{{r}^{2} (x^2-1) {x}^{8}\Big\}}^{-1}\nonumber\\
& - \tfrac{1}{243}\delta\nu''' r \Big\{176\ell {r}^{6} {x}^{2}-56\ell {r}^{6}-351{x}^{8}+108{x}^{6}\Big\} {x}^{-8}\nonumber\\
&+\tfrac{1}{243}\delta\lambda'' \Big\{330\ell {r}^{6} {x}^{4}-328\ell {r}^{6} {x}^{2}+20\ell {r}^{6}-567{x}^{10}+540{x}^{8}-27{x}^{6}\Big\} {\Big\{(x^2-1) {x}^{8}\Big\}}^{-1} \nonumber\\
&- \tfrac{1}{243}\delta\nu'''' {r}^{2} \Big\{14\ell {r}^{6}-27{x}^{6}\Big\} (x^2-1) {x}^{-8}\nonumber\\
&+\tfrac{1}{243}\delta\lambda''' r \Big\{34\ell {r}^{6} {x}^{2}-20\ell {r}^{6}-54{x}^{8}+27{x}^{6}\Big\} {x}^{-8} \,,
\end{align}
and
\begin{align} \label{eq:case1:eom2}
P {r}^{-2}&=- \tfrac{1}{243}\delta\nu'' \Big\{360\ell {r}^{6} {x}^{4}-256\ell {r}^{6} {x}^{2}-4\ell {r}^{6}-486{x}^{10}+378{x}^{8}-27{x}^{6}\Big\} {r}^{-2} {x}^{-8} \nonumber\\
&- \tfrac{1}{243}\delta\nu' \Big\{40\ell {r}^{6} {x}^{4}+22\ell {r}^{6} {x}^{2}+4\ell {r}^{6}+351{x}^{10}-135{x}^{8}+27{x}^{6}\Big\} {r}^{-3} {x}^{-8}\nonumber\\
&+\tfrac{1}{243}\delta\lambda' \Big\{536\ell {r}^{6} {x}^{4}-170\ell {r}^{6} {x}^{2}-48\ell {r}^{6}-837{x}^{10}+351{x}^{8}+81{x}^{6}\Big\} {r}^{-3} {x}^{-8} \nonumber\\
&- \tfrac{2}{243}\delta\lambda \Big\{216\ell {r}^{6} {x}^{4}-66\ell {r}^{6} {x}^{2}-32\ell {r}^{6}-729{x}^{10}+162{x}^{8}+27{x}^{6}\Big\} {r}^{-4} {x}^{-8} \nonumber\\
&- \tfrac{1}{486}\delta\nu'''  \Big\{256\ell {r}^{6} {x}^{2}-76\ell {r}^{6}-378{x}^{8}+135{x}^{6}\Big\}(x^2-1) r^{-1} {x}^{-8}\nonumber\\
&+\tfrac{1}{243}\delta\lambda'' \Big\{330\ell {r}^{6} {x}^{4}-326\ell {r}^{6} {x}^{2}+16\ell {r}^{6}-567{x}^{10}+594{x}^{8}-54{x}^{6}\Big\} {r}^{-2} {x}^{-8} \nonumber\\
&- \tfrac{1}{486}\delta\nu'''' \Big\{20\ell {r}^{6}-27{x}^{6}\Big\} {(x^2-1)}^{2} {x}^{-8}\nonumber\\
&+\tfrac{1}{486}\delta\lambda'''  \Big\{68\ell {r}^{6} {x}^{2}-48\ell {r}^{6}-108{x}^{8}+81{x}^{6}\Big\}(x^2-1) r^{-1} {x}^{-8}\,.
\end{align}
\end{widetext}
A qualitative analysis of the system described by equations \eqref{eq:case1:eom1} and \eqref{eq:case1:eom2} reveals that for any non-zero value of $\ell$, the divergence cannot be resolved as $x$ approaches zero. This issue similarly arises when attempting to set higher-order derivative terms to zero. Consequently, this limitation eliminates the possibility of realizing the theory as expressed in \eqref{eq:case1:f}.

Furthermore, a similar examination of the linearized system indicates that the same critical qualitative behavior persists when substituting \eqref{eq:case1:f} with alternative profiles suggested by (\cite{a}) to either validate or circumvent a cosmological no-go theorem associated with Ricci-inverse theories. These profiles include:
\begin{equation}
f=R+\frac{\alpha}{A},\quad f=R+\alpha R^2 A,\quad f=R + \alpha R e^{-\beta (RA)^2}\,,
\end{equation}
where $\alpha$ and $\beta$ are dimensionless constants. This analysis confirms that none of these models serve as viable Ricci-inverse candidates for sustaining a stable configuration of static spherically symmetric matter distributions.


\section{Circumventing the no-go theorem}
\label{sec:circnogo}

We examine strategies to circumvent the no-go theorem presented in section \ref{sec:nogo}. Our analysis begins by observing that if \( f^{(0)}_A \sim x^{6} \) and \( f^{(0)}_{AA} \sim x^{8} \), then all divergences are eliminated from equations \eqref{eq:pert1} and \eqref{eq:pert2} as \( x \) approaches zero. Hence, given that \( A^{(0)} \sim x^{-2} \), we propose the following model:
\begin{equation}\label{eq:case2:f}
f(R,A) = R + \frac{k}{A^{2}} - 2\Lambda\,,
\end{equation}
where \( k \) and \( \Lambda \) are constant parameters.

The background equation \eqref{eq:dsEOM}, when solved with respect to the parameter $\Lambda$, produces the following result
\begin{equation}\label{eq:case2:L}
\Lambda = 3{x}^{2} {r}^{-2}\,.
\end{equation}
\begin{widetext}
The computation of \eqref{eq:case2:L} within the equations presented in Eqs. \eqref{eq:pert1} and \eqref{eq:pert2} results, as anticipated, in two perturbed equations of motion that are free from divergence:
\begin{align}\label{eq:nodiv1}
\varepsilon {(1-{x}^{2})}^{-1}&=\tfrac{1}{128}\delta\nu'' k \Big\{58{x}^{2}-33\Big\} {x}^{2} {r}^{-2}{(x^2-1)}^{-1}+\tfrac{3}{64}\delta\nu' k \Big\{2{x}^{2}+1\Big\} {x}^{2} {r}^{-3}{(x^2-1)}^{-1}\nonumber\\
&-\tfrac{1}{32}\delta\lambda' \Big\{13k {x}^{4}-2k {x}^{2}+k-32{r}^{2} {x}^{2}+32{r}^{2}\Big\} {r}^{-3} {(x^2-1)}^{-1}+\tfrac{1}{128}\delta\lambda''' k \Big\{{x}^{2}-2\Big\} {r}^{-1}\nonumber\\
&-\tfrac{1}{64}\delta\lambda \Big\{k {x}^{2}-2k-192{r}^{2} {x}^{2}+64{r}^{2}\Big\} {r}^{-4} {(x^2-1)}^{-1}+\tfrac{1}{32}\delta\nu''' k \Big\{4x^2-1\Big\} {r}^{-1}\nonumber\\
&- \tfrac{1}{128}\delta\lambda'' k \Big\{3{x}^{4}+4{x}^{2}-2\Big\} {r}^{-2} {(x^2-1)}^{-1}+\tfrac{1}{128}\delta\nu'''' k \Big\{x^2-1\Big\}\,,
\end{align}
and
\begin{align}\label{eq:nodiv2}
P {r}^{-2}&= - \tfrac{1}{64}\delta\nu'' \Big\{10k {x}^{2}-5k+32{r}^{2} {x}^{2}-32{r}^{2}\Big\} {r}^{-4} +\tfrac{1}{64}\delta\lambda \Big\{-15k {x}^{2}+8k+192{r}^{2} {x}^{2}\Big\} {r}^{-6} \nonumber\\
&- \tfrac{1}{128}\delta\nu' \Big\{-20k {x}^{4}+7k {x}^{2}+10k+256{r}^{2} {x}^{2}-64{r}^{2}\Big\} {r}^{-5}- \tfrac{1}{128}\delta\nu'''' k \Big\{x^2-1\Big\}^{2} {r}^{-2}\nonumber\\
&+\tfrac{1}{128}\delta\lambda' \Big\{-52k {x}^{4}+55k {x}^{2}+128{r}^{2} {x}^{2}-64{r}^{2}\Big\} {r}^{-5}+\tfrac{1}{128}\delta\lambda''' k \Big\{x^2-1\Big\} {x}^{2} {r}^{-3} \nonumber\\
&- \tfrac{1}{128}\delta\nu''' k \Big\{(x^2-1)(8{x}^{2}+1)\Big\} {r}^{-3} - \tfrac{1}{128}\delta\lambda'' k \Big\{3{x}^{4}-13{x}^{2}+8\Big\} {r}^{-4} \,.
\end{align}
\end{widetext}
By design, applying the Sub-Horizon limit as $x\rightarrow 0$ yields
\begin{align}\label{eq:ss1}
\varepsilon&= \tfrac{1}{32}\delta\lambda' \Big\{k+32{r}^{2}\Big\} {r}^{-3} - \tfrac{1}{32}\delta\lambda  \Big\{k-32{r}^{2}\Big\} {r}^{-4}\\& - \tfrac{1}{32}\delta\nu''' k {r}^{-1} - \tfrac{1}{64}\delta\lambda'' k {r}^{-2} - \tfrac{1}{128}\delta\nu'''' k - \tfrac{1}{64}\delta\lambda''' k {r}^{-1}\nonumber \,,
\end{align}
and
\begin{align}\label{eq:ss2}
P {r}^{-2}&=
\tfrac{1}{64}\delta\nu'' \Big\{5k+32{r}^{2}\Big\} {r}^{-4} - \tfrac{1}{64}\delta\nu' \Big\{5k-32{r}^{2}\Big\} {r}^{-5} \nonumber\\&- \tfrac{1}{2}\delta\lambda' {r}^{-3}+\tfrac{1}{8}\delta\lambda k {r}^{-6}+\tfrac{1}{128}\delta\nu''' k {r}^{-3}\nonumber\\& - \tfrac{1}{16}\delta\lambda'' k {r}^{-4} - \tfrac{1}{128}\delta\nu'''' k {r}^{-2}\,.
\end{align}
This finding substantiates the notion that ghosts continue to exist, and no value for \( k \neq 0 \) appears to alleviate their presence.

In the subsequent analysis, we propose a mechanism for constructing ghost-free models in the context of perturbed Ricci-inverse equations \eqref{eq:pert1} and \eqref{eq:pert2}. By employing a methodology analogous to that utilized in the prior case \eqref{eq:case2:f}, it is necessary to demonstrate that the divergence-free characteristics observed in equations \eqref{eq:ss1} and \eqref{eq:ss2} can be replicated when we define the function as follows:
\begin{equation}\label{eq:prof}
f(R,A)=R+kR^{4+i}A^{2+i}-2\Lambda\,,\qquad i \in \mathbb{Z}\,.
\end{equation}

Building upon this premise, since equations \eqref{eq:pert1} and \eqref{eq:pert2} are linear in terms of \(f(R,A)\) and its derivatives, one can formulate a divergence-free model by considering a parameterized linear combination of functionally independent profiles such as \eqref{eq:prof}. This approach allows for the potential avoidance of ghost instabilities by appropriately tuning the free parameters to ensure that higher-order derivatives in \(\delta\nu\) and \(\delta\lambda\) are rendered zero.

To illustrate this concept, we present a straightforward example that exemplifies the aforementioned process. Consider the model defined by:
\begin{equation}\label{eq:proffinal}
f(R,A)= R+\frac{\ell_1}{6} \frac{1}{A^2}+\ell_2 \frac{R}{A}+\frac{\ell_3}{12}R^3A+\ell_4\frac{1}{RA^3}-2\Lambda\,,
\end{equation}
where \(\ell_1, \ell_2, \ell_3, \ell_4\) are constant parameters. Referring back to equations \eqref{eq:pert1} and \eqref{eq:pert2}, we compute the profile given by \eqref{eq:proffinal} and substitute it into the background equation \eqref{eq:dsEOM}. Owing to their divergence-free nature, the resulting Sub-Horizon Weak-Field equations are derived accordingly.
\begin{align}\label{eq:A}
\varepsilon&= \tfrac{1}{1536}\delta\lambda' \Big\{8\ell_1 +1536{r}^{2}+16384+9\ell_4\Big\} {r}^{-3} \nonumber\\&- \tfrac{1}{1536}\delta\lambda \Big\{8\ell_1 -1536{r}^{2}+16384\ell_3+9\ell_4\Big\} {r}^{-4} \nonumber\\
&- \tfrac{1}{192}\delta\nu''' \Big\{96\ell_2 +\ell_1 -4096\ell_3\Big\} {r}^{-1} \nonumber\\&- \tfrac{1}{3072}\delta\lambda'' \Big\{8\ell_1 +16384\ell_3+9\ell_4\Big\} {r}^{-2}\nonumber\\
&-\tfrac{1}{768}\delta\nu'''' \Big\{  96\ell_2  +\ell_1 -4096\ell_3\Big\}\nonumber\\&-\tfrac{1}{3072}\delta\lambda''' \Big\{8\ell_1 +16384\ell_3+9\ell_4\Big\} {r}^{-1} \,,
\end{align}
and
\begin{align}\label{eq:B}
P &{r}^{-2}= \tfrac{1}{1536}\delta\nu'' \Big\{ 768\ell_2 +20\ell_1 +768{r}^{2}+16384\ell_3+15\ell_4\Big\} {r}^{-4} \nonumber\\
&- \tfrac{1}{1536}\delta\nu' \Big\{ 768\ell_2 +20\ell_1 -768{r}^{2}+16384\ell_3+15\ell_4\Big\} {r}^{-5}\nonumber\\
&+\tfrac{1}{3072}\delta\lambda' \Big\{ 768\ell_2 -1536{r}^{2}-49152\ell_3-9\ell_4\Big\} {r}^{-5}\nonumber\\&+\tfrac{1}{768}\delta\lambda \Big\{ 384\ell_2 +16\ell_1 +32768\ell_3+15\ell_4\Big\} {r}^{-6}\nonumber\\
&+\tfrac{1}{3072}\delta\nu''' \Big\{ 768\ell_2 +4\ell_1 -16384\ell_3-3\ell_4\Big\} {r}^{-3}\nonumber\\&-\tfrac{1}{3072}\delta\lambda'' \Big\{ 1536\ell_2 +32\ell_1 +16384\ell_3+21\ell_4\Big\} {r}^{-4} \nonumber\\
&- \tfrac{1}{6144}\delta\nu''''  \Big\{ 8\ell_1 +16384\ell_3+9\ell_4\Big\} {r}^{-2}\nonumber\\& - \tfrac{1}{2048}\delta\lambda''' \Big\{ 256\ell_2 -16384\ell_3-3\ell_4\Big\} {r}^{-3} \,.
\end{align}
It is observed that the higher-order derivative terms associated with $\delta\nu''''$, $\delta\nu'''$, $\delta\lambda''''$, and $\delta\lambda'''$ are now influenced by the coefficients $\ell_1$, $\ell_2$, $\ell_3$, and $\ell_3$. Consequently, it becomes straightforward to determine that these higher-order derivatives vanish in both equations under the condition that:\begin{subequations}
\begin{empheq}[left=\empheqlbrace]{align}
	\, &\ell_1 +96\ell_2  -4096\ell_3=0\,,\\	
        \, &256\ell_2 -16384\ell_3-3\ell_4=0\,,\\	
	\,&4\ell_1 + 768\ell_2  -16384\ell_3-3\ell_4  =0\,.
\end{empheq}
\end{subequations}

The resolution of this system of equations provides 
\begin{equation}
\ell_1(\ell_2) =-\ell_4(\ell_2)=-128\ell_2,
\quad
\ell_3(\ell_2)=-\frac{1}{128}\ell_2,
\end{equation}
that inserted into Eqs. \eqref{eq:A} and \eqref{eq:A} yield
\begin{subequations}
\begin{empheq}[left=\empheqlbrace]{align}
	\, &\delta\lambda'  {r}+\delta\lambda = r^2 \varepsilon\,,\\	
	\,&\delta\nu''r +\delta\nu' - \delta\lambda' = 2P {r}\,.
\end{empheq}
\end{subequations}

By integrating the initial equation and applying the non-relativistic limit \( P \ll \varepsilon \), one can straightforwardly derive the Newtonian potential and curvature perturbations from the definitions provided in \eqref{eq:metricPot}. 

The resulting solutions are:
\[
\frac{d\Phi(r)}{dr} = \frac{M(r)}{r^2} \qquad \frac{d\Psi(r)}{dr} = \frac{M(r)}{r^2}\,.
\]
This outcome appears to be novel within the context of Ricci-inverse theories and aligns perfectly with General Relativity \eqref{eq:metricPotA} at low energy scales.

\section{Discussion and conclusions}
\label{sec:conc}

We have conducted an analysis of the Ricci-inverse modified gravity theory within the framework of a non-relativistic, static, and spherically symmetric cosmic structure situated in a de Sitter cosmology. By considering the Sub-Horizon non-relativistic Weak-Field limit, we discovered that the field equations typically reveal two distinct types of instabilities: (i) divergences that arise when the Sub-Horizon limit is applied, and (ii) the presence of ghosts resulting from terms associated with higher-order derivatives of metric potential perturbations. From this standpoint, we established a novel no-go theorem applicable to small scales. Our findings effectively eliminate all Ricci-inverse models proposed in the literature to address or bypass cosmological and inflationary no-go theorems. Additionally, our investigation prompted a discussion on potential avenues to circumvent the theorem and highlighted a framework for constructing stable models. We demonstrated that these models align completely with the predictions of General Relativity at small scales. Future research will focus on a broader contextualization of our approach and the exploration of new cosmological and astrophysical phenomena related to our findings.



\label{lastpage}
\end{document}